\documentclass[lettersize,journal]{IEEEtran}

\usepackage[T1]{fontenc}
\usepackage[utf8]{inputenc}
\usepackage{textcomp}

\ifCLASSINFOpdf
  \usepackage[pdftex]{graphicx}
\else
  \usepackage[dvips]{graphicx}
\fi

\usepackage[cmex10]{amsmath}
\usepackage{amsfonts}
\usepackage{amssymb}

\usepackage{algorithmic}

\usepackage{array}

\usepackage[hyphens]{url}
\usepackage[breaklinks=true]{hyperref}
\usepackage{cite}
\usepackage{standalone}
\usepackage{pgf, tikz}
\usepackage{pgfplots}
\usepackage{pgfplotstable}
\usepackage{verbatim}
\usetikzlibrary{arrows}
\usetikzlibrary{decorations.text}
\usetikzlibrary{shapes.geometric}
\usetikzlibrary{shapes.arrows}
\usetikzlibrary{shapes}
\usetikzlibrary{positioning}
\usetikzlibrary{shadows}
\usetikzlibrary{patterns}
\usetikzlibrary{arrows.meta}
\pgfplotsset{plot coordinates/math parser=false}
\pgfplotsset{compat=newest}
\pgfplotstableset{use comma,1000 sep=\,}
\pgfkeys{/pgf/number format/.cd,fixed,precision=2}
\usepgfplotslibrary{groupplots}
\definecolor{rwth}   {RGB}{  0  84 159}
\definecolor{rwth-75}{RGB}{ 64 127 183}
\definecolor{rwth-50}{RGB}{142 186 229}
\definecolor{rwth-25}{RGB}{199 221 242}
\definecolor{rwth-10}{RGB}{232 241 250}

\definecolor{black}   {RGB}{  0   0   0}
\definecolor{black-75}{RGB}{100 101 103}
\definecolor{black-50}{RGB}{156 158 159}
\definecolor{black-25}{RGB}{207 209 210}
\definecolor{black-10}{RGB}{236 237 237}

\definecolor{magenta}   {RGB}{227   0 102}
\definecolor{magenta-75}{RGB}{233  96 136}
\definecolor{magenta-50}{RGB}{241 158 177}
\definecolor{magenta-25}{RGB}{249 210 218}
\definecolor{magenta-10}{RGB}{253 238 240}

\definecolor{yellow}   {RGB}{255 237   0}
\definecolor{yellow-75}{RGB}{255 240  85}
\definecolor{yellow-50}{RGB}{255 245 155}
\definecolor{yellow-25}{RGB}{255 250 209}
\definecolor{yellow-10}{RGB}{255 253 238}

\definecolor{petrol}   {RGB}{  0  97 101}
\definecolor{petrol-75}{RGB}{ 45 127 131}
\definecolor{petrol-50}{RGB}{125 164 167}
\definecolor{petrol-25}{RGB}{191 208 209}
\definecolor{petrol-10}{RGB}{230 236 236}

\definecolor{turkis}   {RGB}{  0 152 161}
\definecolor{turkis-75}{RGB}{  0 177 183}
\definecolor{turkis-50}{RGB}{137 204 207}
\definecolor{turkis-25}{RGB}{202 231 231}
\definecolor{turkis-10}{RGB}{235 246 246}

\definecolor{grun}   {RGB}{ 87 171  39}
\definecolor{grun-75}{RGB}{141 192  96}
\definecolor{grun-50}{RGB}{184 214 152}
\definecolor{grun-25}{RGB}{221 235 206}
\definecolor{grun-10}{RGB}{242 247 236}

\definecolor{maigrun}   {RGB}{189 205   0}
\definecolor{maigrun-75}{RGB}{208 217  92}
\definecolor{maigrun-50}{RGB}{224 230 154}
\definecolor{maigrun-25}{RGB}{240 243 208}
\definecolor{maigrun-10}{RGB}{249 250 237}

\definecolor{orange}   {RGB}{246 168   0}
\definecolor{orange-75}{RGB}{250 190  80}
\definecolor{orange-50}{RGB}{253 212 143}
\definecolor{orange-25}{RGB}{254 234 201}
\definecolor{orange-10}{RGB}{255 247 234}

\definecolor{rot}   {RGB}{204   7  30}
\definecolor{rot-75}{RGB}{216  92  65}
\definecolor{rot-50}{RGB}{230 150 121}
\definecolor{rot-25}{RGB}{243 205 187}
\definecolor{rot-10}{RGB}{250 235 227}

\definecolor{bordeaux}   {RGB}{161  16  53}
\definecolor{bordeaux-75}{RGB}{182  82  86}
\definecolor{bordeaux-50}{RGB}{205 139 135}
\definecolor{bordeaux-25}{RGB}{229 197 192}
\definecolor{bordeaux-10}{RGB}{245 232 229}

\definecolor{violett}   {RGB}{ 97  33  88}
\definecolor{violett-75}{RGB}{131  78 117}
\definecolor{violett-50}{RGB}{168 133 158}
\definecolor{violett-25}{RGB}{210 192 205}
\definecolor{violett-10}{RGB}{237 229 234}

\definecolor{lila}   {RGB}{122 111 172}
\definecolor{lila-75}{RGB}{155 145 193}
\definecolor{lila-50}{RGB}{188 181 215}
\definecolor{lila-25}{RGB}{222 218 235}
\definecolor{lila-10}{RGB}{242 240 247}

\usepackage[nolist,nohyperlinks]{acronym}
\acrodef{BESS}[BESS]{Battery Energy Storage System}
\acrodef{ML}[ML]{Machine Learning}
\acrodef{DL}[DL]{Deep Learning}
\acrodef{RL}[RL]{Reinforcement Learning}
\acrodef{IDC}[IDC]{Intraday Continuous}
\acrodef{IDA}[IDA]{Intraday Auction}
\acrodef{DAA}[DAA]{Day Ahead Auction}
\acrodef{WQL}[WQL]{Weighted Quantile Loss}
\acrodef{EPF}[EPF]{Electricity Price Forecasting}
\acrodef{PF}[PF]{Probabilistic Forecasting}
\acrodef{RES}[RES]{Renewable Energy Sources}
\acrodef{TFT}[TFT]{Temporal Fusion Transformer}
\acrodef{SLB}[SLB]{Second Life Battery}
\acrodef{SOC}[SOC]{State-of-Charge}
\acrodef{SOH}[SOH]{State-of-Health}
\acrodef{PWL}[PWL]{Piecewise-linear}
\acrodef{VT}[VT]{Virtual Trade}
\acrodef{BCM}[BCM]{Balancing Capacity Market}
\acrodef{FCR}[FCR]{Frequency Containment Reserve}
\acrodef{aFRR}[aFRR]{automatic Frequency Restoration Reserve}
\acrodef{BTM}[BTM]{Behind-the-Meter}
\acrodef{FTM}[FTM]{Front-of-the-Meter}
\acrodef{MCP}[MCP]{Market Clearing Price}
\acrodef{GCT}[GCT]{Gate Closure Time}
\acrodef{reBAP}[reBAP]{Uniform Imbalance Price}
\acrodef{TSO}[TSO]{Transmission System Operator}
\acrodef{DSO}[DSO]{Distribution System Operator}
\acrodef{OC}[OC]{Opportunity Cost}

\usepackage{siunitx}

\usepackage{orcidlink}

\newcounter{rqcounter}

\begin{document}

\title{Data-Driven Sequential Market Optimization for Front-of-the-Meter Battery Energy Storage Systems}

\author{Steffen~Kortmann\,\orcidlink{0009-0001-2074-773X},~%
Hannah~Sanders,~and~Andreas~Ulbig\,\orcidlink{0000-0001-5834-1842}%
\thanks{This work was co-funded by the European Union in the Horizon Europe i-STENTORE Project under Grant 101096787.
Views and opinions expressed are however those of the author(s) only and do not necessarily reflect those of the European Union or European Climate, Infrastructure and Environment Executive Agency (CINEA). Neither the European Union nor the granting authority can be held responsible for them. \textit{(Corresponding author: S. Kortmann.)}}%
\thanks{S. Kortmann, H. Sanders, and A. Ulbig are with IAEW at RWTH Aachen University, Aachen, Germany, and also with the Center Digital Energy, Fraunhofer FIT, Aachen, Germany (e-mail: s.kortmann@iaew.rwth-aachen.de).}}

% Journal header removed for arXiv preprint; running head shows only the author short-title.
\markboth{Kortmann \MakeLowercase{\textit{et al.}}: Data-Driven Sequential Market Optimization for Front-of-the-Meter BESS}%
{Kortmann \MakeLowercase{\textit{et al.}}: Data-Driven Sequential Market Optimization for Front-of-the-Meter BESS}

\maketitle

\begin{abstract}
The growing integration of \acp{BESS} into electricity markets has highlighted the importance of coordinated participation across energy and balancing services to fully exploit their operational flexibility. However, existing revenue-stacking models often simplify market sequences and neglect the impact of rolling forecasts, leading to unrealistic scheduling and overestimated revenues.

This paper addresses this gap by introducing a sequential, data-driven optimization framework for \ac{FTM} \acp{BESS} that mirrors actual market operations. The framework explicitly models market mechanisms and its respective \ac{GCT} across \ac{FCR}, \ac{aFRR}, \ac{DAA}, and \ac{IDA} markets. Each market stage optimizes expected revenue over all remaining markets using updated price forecasts while maintaining feasibility within both technical and regulatory limits.

A key contribution is the opportunity-cost-based bidding strategy, which endogenously derives market-consistent bid prices and quantities from residual capacity optimization.

Validation for a representative operating day demonstrates that the framework yields consistent and feasible schedules, effectively adapts to updated forecasts, and minimizes deviations between planned and realized dispatch, thereby enhancing the realism and profitability of BESS operation.
\end{abstract}


% % Use this to place sponsorships

% \thanksto{\noindent Submitted to the 24th Power Systems Computation Conference (PSCC 2026).}
\section{Introduction}
\subsection{Motivation}

Germany’s energy transition is accelerating through the large-scale deployment of wind and solar power.  
While these sources are vital for decarbonization, their intermittency introduces significant temporal and spatial volatility, complicating the real-time balance between generation and demand \cite{ec_2050_long_term_strategy}.  
\ac{BESS} are increasingly recognized as a key flexibility resource to manage this variability.  
They support frequency and voltage regulation, mitigate renewable curtailment, and enable more efficient use of excess generation \cite{dnv2024_flexibility_german_power_grid}.  

Installed \ac{BESS} capacity in Germany grew by nearly one-third in 2024, reaching 1.8~GWh, with an additional 3.7~GWh planned by 2027 \cite{reuters2024_german_power_grid_battery_capacity_up}.  
% At the same time, \acp{TSO} have received more than 650 connection requests for large-scale \ac{BESS}, corresponding to a total prospective capacity of approximately 226~GW \cite{enkhardt2025german_grid_226gw}.  
% This unprecedented development emphasizes the need for reliable operational and economic strategies for large \ac{FTM} \ac{BESS}.
The growing number of market participants intensifies competition in ancillary service and energy markets.  
For storage operators, profitability depends on participating across multiple markets—such as \ac{FCR}, \ac{aFRR}, and \ac{DAA} or \ac{IDA} energy trading—while adhering to each market’s specific \ac{GCT} and technical constraints.  
However, most existing optimization approaches assume idealized, simultaneous bidding across markets and do not reflect the sequential and data-driven decision process required in practice.  
Consequently, operators lack transparent methods to determine feasible bid quantities and prices consistent with real market timing and updated forecasts.

\subsection{Related Work}
\label{subsec:related_work}

\begin{table}[t]
\centering
\caption{Comparison of representative \ac{BESS} revenue-stacking models. 
\textbf{(X)} marks data-driven market coupling by rolling forecasts.}
\label{table_literature}
\resizebox{\columnwidth}{!}{%
\begin{tabular}{|l|c|c|c|c|l|l|}
\hline
\textbf{Reference} & \multicolumn{4}{c|}{\textbf{Energy \& Balancing Markets}} & \textbf{Scheduling Approach} & \textbf{Bidding Strategy} \\
\cline{2-5}
 & \textbf{FCR} & \textbf{aFRR} & \textbf{DA} & \textbf{IDA} & & \\
\hline
\cite{englbergerOptimizedEnergyManagement2021} & X & - & X & – & Single-Stage Optimization & N/A \\
\hline
\cite{mohamed_stacked_2023} & X & X & X & X & Single-Stage Optimization & N/A \\
\hline
\cite{cortesM5UseOptimizationFramework2024} & X & X & - & - & Single-Stage Optimization & N/A \\
\hline
\cite{baltputnisRobustMarketbasedBattery2024} & X & X & – & X & Robust Optimization & Residual Energy \\
\hline
\cite{cremonciniOptimalParticipationWind2024} & – & X & X & – & Robust Optimization & Fixed Coupling \\
\hline
\cite{martinez-ricoEnergyStorageSizing2021} & – & X & X & – & Single-Stage Optimization & N/A \\
\hline
\cite{paredesVirtualEnergyStorage2024} & - & - & X & - & Single-Stage Optimization & N/A \\
\hline
\cite{paredesRevenueStackingBESSs2024} & – & X & X & – & Single-Stage Optimization & N/A \\
\hline
\cite{tadayonMultiLevelSimulationFramework2025} & X & X & X & – & Sequential Optimization & N/A \\
\hline
\cite{kortmann_using_2025} & - & - & X & X & Sequential Optimization & N/A \\
\hline
\cite{merten_participation_2020} & – & X & – & – & Machine Learning & Statistical \\
\hline
\cite{biggins_trade_2022} & - & X & - & - & Single-Stage Optimization & Learning-Based \\
\hline
\cite{dong_strategic_2021} & – & X & - & – & Reinforcement Learning & Learning-Based \\
\hline
\cite{he_optimal_2016} & – & - & X & – & Single-Stage Optimization & N/A \\
\hline
\cite{spillerStochasticVsMonte2024} & – & – & X & X & Single-Stage Optimization & Stochastic/Monte Carlo \\
\hline
\cite{rancilioBESSAncillaryServices2024} & - & X & - & – & Sequential Optimization & Statistical \\
\hline
\textbf{This work} & (X) & (X) & (X) & (X) & \textbf{Sequential Optimization} & \textbf{Opportunity Cost} \\
\hline
\end{tabular}%
}
\end{table}

Revenue stacking has been widely identified as a key driver of battery profitability, yet most formulations fail to capture the sequential nature of market participation and the impact of rolling forecasts \cite{englbergerOptimizedEnergyManagement2021}.  

Recent frameworks propose multi-market coordination \cite{mohamed_stacked_2023, cortesM5UseOptimizationFramework2024}, robust co-optimization of \ac{FCR} and \ac{aFRR} \cite{baltputnisRobustMarketbasedBattery2024}, and DA–aFRR participation models \cite{cremonciniOptimalParticipationWind2024}, though most assume concurrent bidding, inconsistent with actual sequential market operation.  
Additional studies focus on aFRR/mFRR market participation and sizing \cite{martinez-ricoEnergyStorageSizing2021,paredesVirtualEnergyStorage2024}.
Integrated multi-market formulations spanning wholesale and balancing services introduce uncertainty, which can be addressed using joint-chance-constraint techniques \cite{paredesRevenueStackingBESSs2024}. 
Sequential formulations better reflect real market timing but typically omit bidding strategies, overestimating achievable revenues \cite{tadayonMultiLevelSimulationFramework2025}.
% The work by \cite{tadayonMultiLevelSimulationFramework2025} applies a sequential multi-market formulation, but omits bidding strategies and therefore overestimates the revenue potential by \ac{BESS}.

Data-driven and learning-based approaches have emerged more recently.  
Probabilistic deep-learning forecasts have improved market price prediction \cite{kortmann_using_2025}, while machine-learning-based bidding models estimate aFRR bid acceptance probabilities \cite{merten_participation_2020,biggins_trade_2022}.  
Recent studies employ \ac{RL} for strategic bidding \cite{dong_strategic_2021}; however, such models are sensitive to data distribution shifts and require continual retraining to remain effective.
Optimization-based bidding with regard to regulatory constraints is shown in  \cite{he_optimal_2016}. 
Stochastic and Monte-Carlo-based bidding strategies for sequential markets \cite{spillerStochasticVsMonte2024} further expand the literature.  
In \cite{rancilioBESSAncillaryServices2024}, bidding and control strategies for \ac{BESS} are presented that incorporate forecast models, but solely focus on balancing markets. The bidding is data-driven, but not based on opportunity-cost pricing with regard to the other markets.

To our knowledge, we are the first to provide a unified, data-driven framework that integrates rolling forecasts, opportunity-cost-based bidding, and market-accurate gate-closure timing across day-ahead, intraday, and balancing markets.

\subsection{Contributions} \label{subsec:contributions}
Our work addresses these gaps by introducing a sequential, data-driven optimization framework for \ac{FTM} \ac{BESS} operation.  

Our main contributions are as follows:

\begin{enumerate}
    \item Sequential multi-market formulation: A unified optimization framework that mirrors actual market sequences with regards to \acp{GCT} and product durations for \ac{FCR}, \ac{aFRR}, and energy markets.
    \item Data-driven market coupling: Integration of rolling price forecasts into the optimization, enabling adaptive re-optimization at each market stage with price information from already cleared markets.
    \item Opportunity-cost-based bidding: Derivation of feasible bid volumes and minimum bid prices from residual capacity optimization, ensuring economic consistency and avoiding infeasible commitments.
    \item Validation under realistic operation: Demonstration of framework consistency through a single day, confirming alignment between day-ahead and dispatch schedules.
\end{enumerate}

% This study focuses exclusively on \ac{FTM} applications, excluding \ac{BTM} use cases such as residential or commercial storage.  
The central research question guiding our work is therefore:  
\textit{(i) How can a data-driven optimization algorithm be designed to schedule BESS across power exchange and balancing markets in line with realistic market deciion-making, and (ii) what are the resulting implications for the operational behavior of the batteries?}
% \RQ{rq:optimization}{\textbf{\textit{How can a data-driven optimization algorithm be designed to schedule BESS across power exchange and balancing markets in line with realistic market decision-making, and what are the resulting implications for the operational behavior of the batteries?}}}
% Addressing this question requires the integration of market forecasting, operational constraints, and sequential decision-making within one coherent optimization framework.

\section{Methodology}\label{sec:methodology}

The scheduling of \ac{BESS} is performed in consecutive market stages, where both technical and economic constraints are continuously taken into account. To reflect these requirements, the dispatch optimization is structured sequentially, consisting of day-ahead planning and subsequent intraday adjustments. 
% In each optimization step, constraints arising from market rules and from the physical properties of the battery must be respected.

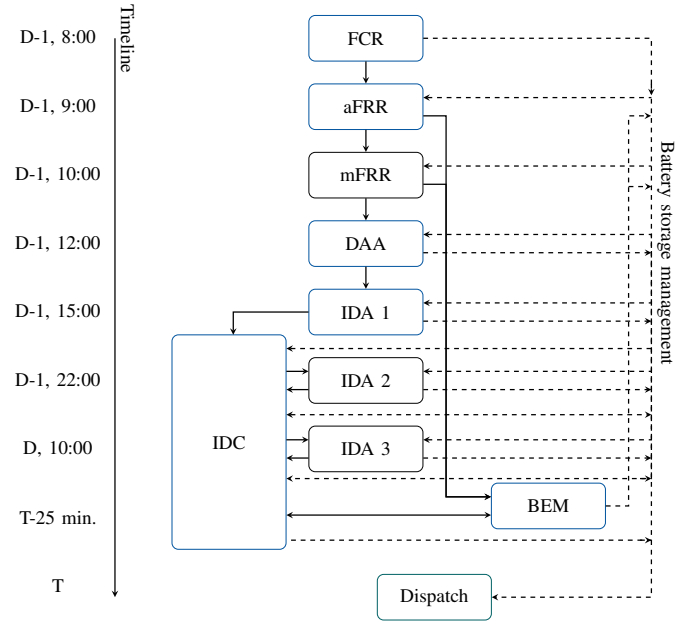
\begin{figure}[ht]
    \centering
    \resizebox{\columnwidth}{!}{%
\begin{tikzpicture}[node distance=1.5cm, font=\large]
    \tikzstyle{block} = [rectangle, rounded corners, draw=black, text centered, minimum width=2.5cm, minimum height=1cm]
    \tikzstyle{blockIDC} = [rectangle, rounded corners, draw=black, text centered, minimum width=2.5cm, minimum height=4.7cm]
    \tikzstyle{arrow} = [thick,->,>=stealth]
    \tikzstyle{dashedarrow} = [thick,->,>=stealth, dashed]

 % Blocks
    \node (fcr) [block, draw=rwth] {FCR};       %8:00, aFRR, DA, IDA
    %\node (fcr) [block, below of=start] {FCR-Ergebnisse};   %8:30 
    \node (afrr) [block, below of=fcr, draw=rwth] {aFRR};    % (9:00 DA, IDA1-3, IDC)
    \node (mfrr) [block, below of=afrr] {mFRR};
    \node (daa) [block, below of=mfrr, draw=rwth] {DAA};
    \node (ida1) [block, below of=daa, draw=rwth] {IDA 1};
    \node (ida2) [block, below of=ida1] {IDA 2};
    \node (ida3) [block, below of=ida2] {IDA 3};
    \node (idc) [blockIDC, below of=ida3, xshift=-3cm, yshift=1.65cm, draw=rwth] {IDC};
    \node (ram) [block, right of=ida3, xshift=2.5cm, yshift=-1.25cm, draw=rwth] {BEM};
    \node (handlung) [block, below of=ram, xshift=-2.5cm, yshift=-0.5cm, draw=petrol] {Dispatch};

    % Arrows unten
    %\draw [arrow] (start) -- (fcr);
    \draw [arrow] (fcr) -- (afrr);    %\draw [arrow] (fcr) -- node[left] {Rest} (afrr);
    \draw [arrow] (afrr) -- (mfrr);
    \draw [arrow] (mfrr) -- (daa);
    \draw [arrow] (daa) -- (ida1);
    %\draw [arrow] (ida1) -- (ida2);
    %\draw [arrow] (ida2) -- (ida3);
    \draw [arrow] (ida1.west) -- ++(-1.65cm,-0) -- ++ (0cm,-0.5);
    \draw [arrow, <-] ($(ida2.west)+ (0,0.2)$) --  ($(idc.east)+ (0,1.55)$);
    \draw [arrow] ($(ida2.west)+ (0,-0.2)$) --  ($(idc.east)+ (0,1.15)$);
    \draw [arrow, <-] ($(ida3.west)+ (0,0.2)$) -- ($(idc.east)+ (0,0.05)$);
    \draw [arrow] ($(ida3.west)+ (0,-0.2)$) -- ($(idc.east)+ (0,-0.35)$);

    % Arrows RAM
    \draw [arrow] ($(afrr.east)+ (0,-0.2)$) -- ++(0.5cm,0) |- ($(ram.west)+ (0,0.2)$);
    \draw [dashedarrow] (ram.east) -- ++(0.5cm,0) |- (6.25cm,-1.7);
    \draw [arrow, <->] ($(idc.east)+ (0,-1.6)$) -- ($(ram.west)+ (0,-0.2)$);

    %\draw [arrow] ($(mfrr.east)+ (0,-0.2)$) -- ($(mfrr.east)+ (0.5,-0.2)$);
    \draw [arrow] ($(mfrr.east)+ (0,-0.2)$) -- ++(0.5cm,0) |- ($(ram.west)+ (0,0.2)$);
    \draw [dashedarrow] ($(ram.east)+ (0.5,7)$) -- ++(0.5cm,0);

    % Seitliche Pfeile zum Hauptstrang
    \draw [dashedarrow] (fcr.east) -- ++(5cm,0) |- ($(idc.east)+ (0,-0.8)$);
    %\draw [dashedarrow] ($(afrr.east)+ (0,-0.2)$) -- ++(5cm,0) |- (idc.east);
    %\draw [dashedarrow] ($(mfrr.east)+ (0,-0.2)$) -- ($(mfrr.east)+ (5,-0.2)$);
    \draw [dashedarrow] ($(daa.east)+ (0,-0.2)$) -- ($(daa.east)+ (5,-0.2)$) node [rotate=-90, above] {Battery storage management}; %Speichermanagement Nebenbedingungen
    \draw [dashedarrow] ($(ida1.east)+ (0,-0.2)$) -- ($(ida1.east)+ (5,-0.2)$);
    \draw [dashedarrow] ($(ida2.east)+ (0,-0.2)$) -- ($(ida2.east)+ (5,-0.2)$);
    \draw [dashedarrow] ($(ida3.east)+ (0,-0.2)$) -- ($(ida3.east)+ (5,-0.2)$);
    
    \draw [dashedarrow, <-] ($(idc.east)+ (0,2.05)$) -- ($(idc.east)+ (8,2.05)$);
    \draw [dashedarrow, <->] ($(idc.east)+ (0,0.6)$) -- ($(idc.east)+ (8,0.6)$);
    \draw [dashedarrow] ($(idc.east)+ (7.75,-0.8)$) -- ++(0.25cm,0);

    \draw [dashedarrow] ($(fcr.east)+ (5,-1)$) -- ($(fcr.east)+ (5,-1.25)$);
    \draw [dashedarrow] ($(idc.east)+ (7.75,-2.15)$) -- ++(0.25,0);

    % Seitliche Pfeile vom Hauptstrang weg
    %\draw [dashedarrow] (fcr.east) -- ++(5cm,0) |- (idc.east);
    \draw [dashedarrow, <-] ($(afrr.east)+ (0,0.2)$) -- ($(afrr.east)+ (5,0.2)$);
    \draw [dashedarrow, <-] ($(mfrr.east)+ (0,0.2)$) -- ($(mfrr.east)+ (5,0.2)$);
    \draw [dashedarrow, <-] ($(daa.east)+ (0,0.2)$) -- ++(5cm,0) |- ($(idc.east)+ (0,-2.15)$);
    \draw [dashedarrow, <-] ($(ida1.east)+ (0,0.2)$) -- ($(ida1.east)+ (5,0.2)$);
    \draw [dashedarrow, <-] ($(ida2.east)+ (0,0.2)$) -- ($(ida2.east)+ (5,0.2)$);
    \draw [dashedarrow, <-] ($(ida3.east)+ (0,0.2)$) -- ($(ida3.east)+ (5,0.2)$);

    \draw [dashedarrow] ($(ida3.east)+ (5,-2)$) -- ++ (0,-1) |- (handlung.east);

    %Zeitstrahl
    \draw[arrow] (-5.5,0) node [rotate=-90, above] {Timeline}--++ (0, -12.25);
    \node (d1_8) [ left of = fcr, xshift=-5.25cm] {D-1, 8:00};       %8:00, aFRR, DA, IDA
    \node (d1_9) [ below of=d1_8] {D-1, 9:00};    % (9:00 DA, IDA1-3, IDC)
    \node (d1_10) [ below of=d1_9] {D-1, 10:00};
    \node (d1_12) [ below of=d1_10] {D-1, 12:00};
    \node (d1_15) [ below of=d1_12] {D-1, 15:00};
    \node (d1_22) [ below of=d1_15] {D-1, 22:00};
    \node (d_10) [ below of=d1_22] {D, 10:00};
    \node (t_25) [  below of=d_10] {T-25 min.};
    %\node (t_5) [block, below of=t_25] {T-5 min.};
    \node (t) [ below of=t_25] {T};

\end{tikzpicture}%
}
\caption{Flow chart of optimization framework (considered markets in \textcolor{rwth}{blue}).}
\label{fig:Ablauf}
\end{figure}

Figure \ref{fig:Ablauf} shows the basic idea of the chronological sequence of the sequential decision-making problem of optimization. Basically, the current state of knowledge, such as the foreseeable technical status or price forecasts, is taken at a specific gate closure time. Rolling market forecasts are updated at each gate closure, ensuring that every optimization step reflects the most recent information from prior market results and forecast revisions. On this basis, optimizations are carried out for the market that is currently closing, so that one or more bids are submitted. The award is then simulated. It is assumed here that the bid for the battery itself does not influence the \ac{MCP}. Subsequently, the obligations arising from a possible awarded bid, as well as the price structure of the closed market, are available for subsequent decisions.
\section{Modeling}\label{sec:modeling}

In each individual point in time of the sequential optimization, there are general constraints that must be met. These can be divided into those constraints that arise from the battery and those from markets or market rules, as seen in Fig.~\ref{fig:fcu}.

\usetikzlibrary{calc}
\usetikzlibrary{patterns}

% \tikzstyle{every node}=[font=\small]
\tikzstyle{every path}=[line width=0.8pt,line cap=round,line join=round]

\begin{figure} [htbp]
\centering
\resizebox{\columnwidth}{!}{%
\begin{tikzpicture}[font=\large]
    %\centering

    % Block Positionen
    \coordinate (Battery) at (0,0);
    \coordinate (DC_AC) at (4,0);

    % Rahmen um das System
    \draw[dashed, rwth, thick] (-1.2,-1.8) rectangle (5.6,1.8);
    \node[rwth] at (2.25,-2.1) {\textbf{Battery constraints}};
    \node[black] at (2.25,1.5) {AC coupled BESS};
    
%    \draw[dashed, magenta, thick] (9,-1.8) rectangle (13,1.8);
    \node[magenta] at (10.5,-2.1) {\textbf{Market constraints}};
    
    % Batterie
    \node[draw, ultra thick, minimum width=1.2cm, minimum height=1.5cm, anchor=center] at (Battery) {};
%    \draw[line width=1pt] (-0.5,0.5) -- (0.5,-0.5); % Blitzsymbol
    \draw (0, 0.5) -- (0, 0.09);
    \draw (0, -0.09) -- (0, -0.5);
    \draw [line width=1.5pt] (0.3, 0.09) -- (-0.3, 0.09);
    \draw [line width=2pt] (0.2, -0.09) -- (-0.2, -0.09);
    \node at ($(Battery)+(0,-1.1)$) {BESS};

    % DC-AC Wandler
    \draw[ultra thick] (DC_AC) ++(-0.8,-0.8) rectangle ++(1.6,1.6);
    \draw[white, fill=black] (DC_AC) ++(-0.8,-0.8) rectangle ++(1.6,1.6);
    \draw[white, line width=0.8pt] (DC_AC) ++(-0.8,-0.8) -- ++(1.6,1.6);
    \node[white] at ($(DC_AC)+(0,0.5)$) {=};
    \node[white] at ($(DC_AC)+(0,-0.5)$) {$\sim$};
    \node at ($(DC_AC)+(0,-1.1)$) {DC-AC inverter};

    % Verbindungslinien
    \draw[dashed, thick] (0.6,0) -- ($(DC_AC)+(-0.8,0)$);
    \draw[dashed, thick] ($(DC_AC)+(0.8,0)$) -- ++(3,0); % weitere Verbindung angedeutet

    % p_charge/discharge bat
    \node[anchor=west] at ($(Battery)+(1,0.6)$) (pBc){$p^{\text{bat}}_{\text{dc\_charge}}$};
    \node[anchor=west] at ($(Battery)+(1,-0.6)$) (pBd) {$p^{\text{bat}}_{\text{dc\_discharge}}$};
    \draw[<-] (pBc.south) ++(-0.3,-0.08) -- ++(0.6,0); % kurzer Pfeil über dem Text
    \draw[->] (pBd.north) ++(-0.3,0.08) -- ++(0.6,0);

    % p_charge/discharge markets
    \node[anchor=west] at ($(DC_AC)+(1.8,0.6)$) (pMc){$p^{\text{markets}}_{\text{ac\_charge}}$};
    \node[anchor=west] at ($(DC_AC)+(1.8,-0.6)$) (pMd) {$p^{\text{markets}}_{\text{ac\_discharge}}$};
    \draw[<-] (pMc.south) ++(-0.3,-0.08) -- ++(0.6,0); % kurzer Pfeil über dem Text
    \draw[->] (pMd.north) ++(-0.3,0.08) -- ++(0.6,0);

    %Vertikale Linie
    \draw[dashed, thick] ($(DC_AC)+(3.8,1.77)$) -- ($(DC_AC)+(3.8,-2)$);

    % --- move PCC side right ---
% Grid-Knoten
\node[draw, ultra thick, minimum width=1.5cm, minimum height=1.5cm,
      pattern=crosshatch, pattern color=black-50]
      at ($(DC_AC)+(4.3,2.5)$) {Grid};

\node at ($(DC_AC)+(4.7,0)$) {PCC};

\node[anchor=west] at ($(DC_AC)+(5.5,1.5)$) {Capacity Markets};
\node[anchor=west, draw=magenta, dashed, minimum width=1.2cm]
      at ($(DC_AC)+(6.0,1)$) {FCR};
\node[anchor=west, draw=magenta, dashed, minimum width=1.2cm]
      at ($(DC_AC)+(6.0,0.4)$) {aFRR};

\node[anchor=west] at ($(DC_AC)+(5.5,-0.4)$) {Power Exchange Markets};
\node[anchor=west, draw=magenta, dashed, minimum width=1.2cm]
      at ($(DC_AC)+(6.0,-0.9)$) {DAA};
\node[anchor=west, draw=magenta, dashed, minimum width=1.2cm]
      at ($(DC_AC)+(6.0,-1.5)$) {IDA1};

\draw [decorate, decoration={brace, mirror, amplitude=10pt}]
    ($(DC_AC)+(5.6,1.7)$) -- ($(DC_AC)+(5.6,-1.7)$);

\end{tikzpicture}%
}

%\end{center}
\caption{Split in modeling framework between battery and market constraints}
    \label{fig:fcu}
\end{figure}
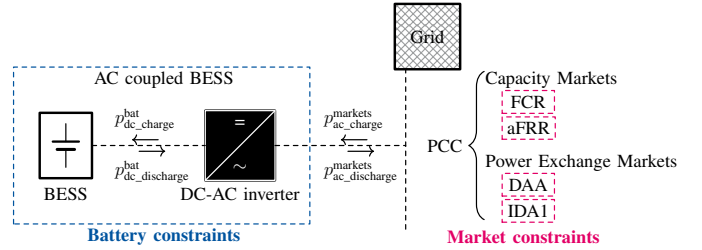

\subsubsection{Battery Constraints}

This section covers the physical representation of the battery within the optimization model. 
The constraints ensure that charging and discharging processes respect technical limits on power, capacity, efficiencies, and operating logic, while accurately modeling the dynamic evolution of the \ac{SOC} over time.

The dynamic behavior of the battery's \ac{SOC} is described by the energy balance equation that accounts for both charging and discharging processes~\eqref{eq:soc_balance}. 
Hence, the \ac{SOC} at the end of time step~$t$ depends on the \ac{SOC} at the end of the previous step and the battery actions in the current timestep.
%State of Charge (SoC) dynamisch über die Zeit
\begin{equation}
\mathrm{SoC}_t 
= \mathrm{SoC}_{t-1} 
+ \left( \eta_{\mathrm{in}} P^{\mathrm{ch}}_t 
- \frac{P^{\mathrm{dis}}_t}{\eta_{\mathrm{out}}} \right) \cdot \Delta t,
\quad \forall\, t \in \mathcal{T}.
\label{eq:soc_balance}
\end{equation}

To ensure consistent operation over the time horizon under consideration, the initial and final relative states of charge $\mathrm{SoC}_{\mathrm{ini}}$ and $\mathrm{SoC}_{\mathrm{final}}$ are specified externally. 
A small tolerance~$\varepsilon$ on the final \ac{SOC} can be allowed to accelerate the solution process (see ~\eqref{eq:soc_initial_final_eps_rel}). 
Unless otherwise stated, the initial and final relative \ac{SOC} are chosen to be equal.
%Initial and Final SoC
% Initial and terminal SoC with small tolerance (relative, 0–1)
\begin{equation}
\mathrm{SoC}_0 = \mathrm{SoC}_{\mathrm{ini}},
\quad
\bigl|\mathrm{SoC}_T - \mathrm{SoC}_{\mathrm{final}}\bigr| \le \varepsilon.
\label{eq:soc_initial_final_eps_rel}
\end{equation}

Physically, charging and discharging cannot occur simultaneously. 
%\paragraph{Mutual Exclusivity of Charging and Discharging.}
This mutual exclusivity is enforced by constraining the sum of the binary variables $b^{\mathrm{ch}}_t$ and $b^{\mathrm{dis}}_t$ to be at most one for each time step, as shown in \eqref{eq:xor_ch_dis}.
%Laden und Entladen nicht gleichzeitig (XOR)
\begin{equation}
b^{\mathrm{ch}}_t + b^{\mathrm{dis}}_t \;\le\; 1,
\quad \forall\, t \in \mathcal{T}.
\label{eq:xor_ch_dis}
\end{equation}

%\paragraph{Power Limitation.}
In addition, actual charging and discharging powers are only allowed when their respective binary indicators are active. In addition, these powers are restricted by the battery’s maximum rated power, as given in \eqref{eq:power_limiting}.
% Lade- und Entladeleistung nur bei aktivem Status zulassen
\begin{equation}
P^{\mathrm{ch}}_t \leq b^{\mathrm{ch}}_t \cdot P^{\mathrm{bat}}_{\max}
\quad , \quad
P^{\mathrm{dis}}_t \leq b^{\mathrm{dis}}_t \cdot P^{\mathrm{bat}}_{\max},
\quad \forall\, t \in \mathcal{T}.
\label{eq:power_limiting}
\end{equation}

The net battery power $P^{\mathrm{bat}}_t$ is defined as the algebraic sum of charging and discharging power (positive for charging, negative for discharging), see \eqref{eq:pbat_def}. 
Its absolute value is limited by the battery’s maximum power rating, compare \eqref{eq:pbat_bounds}.
% Definition of net battery power
\begin{equation}
P^{\mathrm{bat}}_t \;=\; P^{\mathrm{ch}}_t \;-\; P^{\mathrm{dis}}_t,
\quad \forall\, t \in \mathcal{T}.
\label{eq:pbat_def}
\end{equation}

% Bounds on net battery power
\begin{equation}
-\,P^{\mathrm{bat}}_{\max} \;\le\; P^{\mathrm{bat}}_t \;\le\; P^{\mathrm{bat}}_{\max},
\quad \forall\, t \in \mathcal{T}.
\label{eq:pbat_bounds}
\end{equation}

This basic battery model can be equipped with additional constraints, e.g., with degradation models, which is outside the scope of this work.

\subsubsection{Market Constraints}

To adequately model energy trading in power exchange and reserve markets, various constraints are imposed to reflect operational and market design limitations.

In energy exchange markets $(m \in \mathcal{M}_{ex})$, traded power is restricted to fixed increments $\Delta q_m$ [MW]. 

This is implemented by auxiliary integer variables $z^{\mathrm{buy}}_{m,t}$ and $z^{\mathrm{sell}}_{m,t}$ that scale bid increment, as defined in \eqref{eq:buy_step_constraint}--\eqref{eq:sell_step_constraint}. 
% These constraints ensure that buy and sell quantities are integer multiples of market-specific minimum bid size. 
%step-size
\begin{equation}
q^{\mathrm{buy}}_{m,t} =  \Delta q_m \cdot z^{\mathrm{buy}}_{m,t}
\quad \forall m \in \mathcal{M}_{ex},\; t\in\mathcal{T}
\label{eq:buy_step_constraint}
\end{equation}

\begin{equation}
q^{\mathrm{sell}}_{m,t} =  \Delta q_m \cdot z^{\mathrm{sell}}_{m,t}
\quad \forall m \in \mathcal{M}_{ex},\; t\in\mathcal{T}
\label{eq:sell_step_constraint}
\end{equation}

For balancing reserve markets $(m \in \mathcal{M}_{cap})$, no explicit step-size constraint is required, since decision variables $P^{\mathrm{res}}_{m,t}$ are already integer-valued in MW. 
With a minimum bid increment of 1~MW, integrality alone enforces the step size.
Equations~\eqref{eq:buy_step_constraint}--\eqref{eq:sell_step_constraint} can optionally be disabled to reduce computational effort. 
In that case, available tradable power in the bidding strategy must be rounded down to a multiple of $\Delta q_m$.

In addition, the \ac{SOC} must also reflect potential energy requirement resulting from allocated balancing reserve capacity.
%\paragraph{Reserve-Induced SoC Limits.}
Equation~\eqref{eq:soc_min_due_to_reserve} ensures that sufficient energy is available to deliver positive reserves, while \eqref{eq:soc_max_due_to_reserve} ensures that enough storage headroom is reserved to absorb potential charging needs from negative reserves. 
\begin{equation}
\label{eq:soc_min_due_to_reserve}
\mathrm{SoC}_{t-1} \;\ge\;
\sum_{m \in \mathcal{M}_{pos}}
\left(
\frac{P^{\mathrm{res}}_{m,t}\,\gamma_m}{\eta_{\mathrm{out}}}
\right),
\quad \forall\, t \in \mathcal{T}
\end{equation}

\begin{equation}
\label{eq:soc_max_due_to_reserve}
\mathrm{SoC}_{t-1} \;\le\; C^{\mathrm{bat}} -
\sum_{m \in \mathcal{M}_{neg}}
\left(
P^{\mathrm{res}}_{m,t}\,\gamma_m\,\eta_{\mathrm{in}}
\right),
\quad \forall\, t \in \mathcal{T}
\end{equation}

These constraints represent the blocking of usable storage capacity caused by balancing capacity commitments, scaled by the market-specific energy-to-capacity ratio~$\gamma_m$ [MWh/MW]. 
Due to its symmetric nature, the FCR product belongs to both the set of positive reserve markets~$\mathcal{M}_{pos}$ and the set of negative reserve markets~$\mathcal{M}_{neg}$. 
FCR requires a minimum energy-to-power ratio of $\gamma_{\mathrm{FCR}} = \frac{0.91}{2}~\si{\mega\watt\hour\per\mega\watt}$, whereas FRR products require $\gamma_{\mathrm{FRR}} = 1~\si{\mega\watt\hour\per\mega\watt}$ per direction.

The net sum of all market energy transactions at each time step must respect the physical charging and discharging limits of the battery, while also considering power blocked by balancing reserve commitments. 
%\paragraph{Limits on Trading Based on Battery Power Capacity.}%
This is enforced through \eqref{eq:upper_market_power_limit} and\eqref{eq:lower_market_power_limit}, which set upper and lower bounds on the aggregated market flow.

Batteries participating in balancing reserve markets are subject to additional requirements regarding available power capacity, expressed by the blocked power factor~$\phi_m$ per MW of reserved capacity in the balancing market~$m$. 
For example, the FCR requires $\phi_{\mathrm{FCR}} = 1.25$, meaning that symmetrical provision of $1\,\si{\mega\watt}$ reserves requires $1.25\,\si{\mega\watt}$ of available charging and discharging power. 
The additional $0.25\,\si{\mega\watt}$ is reserved exclusively for intraday recharge management. 
Accordingly, $\phi_{\mathrm{FCR}}$ is set to $1.0$ in the day-ahead and $1.25$ in the intraday optimization. 
For FRR products, $\phi_{\mathrm{FRR}} = 2.0$ applies in the day-ahead (two MW in each direction for every MW marketed), while intraday trading uses $\phi_{\mathrm{FRR}} = 1.0$. 
The power thus released can be traded on the \ac{IDC} while still respecting the \ac{SOC} limits during activation.
\begin{equation}
\sum_{m \in \mathcal{M}_{ex}} 
\left( q^{\mathrm{buy}}_{m,t} - q^{\mathrm{sell}}_{m,t} \right)
\leq 
P^{\mathrm{bat}}_{\max} -
\sum_{r \in \mathcal{M}_{neg}} 
\left( \phi_r \cdot P^{\mathrm{res}}_{r,t} \right)
\label{eq:upper_market_power_limit}
\end{equation}

\begin{equation}
\sum_{m \in \mathcal{M}_{ex}} 
\left( q^{\mathrm{buy}}_{m,t} - q^{\mathrm{sell}}_{m,t} \right)
\geq 
- P^{\mathrm{bat}}_{\max} +
\sum_{r \in \mathcal{M}_{pos}} 
\left( \phi_r \cdot P^{\mathrm{res}}_{r,t} \right)
\label{eq:lower_market_power_limit}
\end{equation}

% \begin{equation}
% \forall\, t \in \mathcal{T} \nonumber
% \end{equation}

Several markets require power trading decisions to remain constant within predefined time blocks. 
%\paragraph{Block Consistency in Market Offers.}
To capture this behavior, \eqref{eq:market_buy_consistency} and~\eqref{eq:market_sell_consistency} enforce consistency for bought and sold power quantities across all time steps that do not mark the end of a block. 
Similarly, \eqref{eq:market_reserve_consistency} ensures constant reserve provision levels within the block structure of each reserve product, as defined by the set~$\mathcal{B}_m$ of block end indices for market~$m$. 
% The block durations for the individual markets are given in Table~\ref{tab:regelreserve} and Table~\ref{tab:power_exchange_markets}.
\begin{equation}
q^{\mathrm{buy}}_{m,t} = q^{\mathrm{buy}}_{m,t+1},
\quad \forall m \in \mathcal{M}_{ex},\; t \in \mathcal{T} \setminus \mathcal{B}_m
\label{eq:market_buy_consistency}
\end{equation}

\begin{equation}
q^{\mathrm{sell}}_{m,t} = q^{\mathrm{sell}}_{m,t+1},
\quad \forall m \in \mathcal{M}_{ex},\; t \in \mathcal{T} \setminus \mathcal{B}_m
\label{eq:market_sell_consistency}
\end{equation}

\begin{equation}
P^{\mathrm{res}}_{m,t} = P^{\mathrm{res}}_{m,t+1},
\quad \forall m \in \mathcal{M}_{cap},\; t \in \mathcal{T} \setminus \mathcal{B}_m
\label{eq:market_reserve_consistency}
\end{equation}

The net flows of the markets must be consistent with the battery (dis-)charging decisions at every time step~$t$.  
%\paragraph{Coupling Between Market Transactions and Battery Operation.}
For discharging, \eqref{eq:discharge_coupling} and\eqref{eq:discharge_coupling_b} implement a match-if-active logic. If the discharge indicator $b^{\mathrm{dis}}_t=1$, then the net exported power $\sum_{m \in \mathcal{M}_{ex}} (q^{\mathrm{sell}}_{m,t} - q^{\mathrm{buy}}_{m,t})$ must be equal to the discharge output $P^{\mathrm{dis}}_t$, up to the slack $s^{\mathrm{dis}}_t$. 
Equation~\eqref{eq:discharge_binary_coupling} adds the general limit that exports cannot exceed the available discharge power. 
Analogously, \eqref{eq:charge_coupling} and~\eqref{eq:charge_coupling_b} enforce that whenever the charging indicator $b^{\mathrm{ch}}_t=1$, the net imported power 
$\sum_{m \in \mathcal{M}_{ex}} (q^{\mathrm{buy}}_{m,t} - q^{\mathrm{sell}}_{m,t})$ 
matches the charging output 
$P^{\mathrm{ch}}_t$ up to the slack $s^{\mathrm{ch}}_t$, 
while \eqref{eq:charge_binary_coupling} ensures that buy and sell quantities remain within the charging limit. 
In the strict formulation, both slack variables are fixed to zero; in the relaxed formulation, they are penalized to maintain feasibility in edge cases such as activation or rounding effects. 
A bilinear coupling of binaries and flows would be conceptually simpler but is avoided in favor of this formulation, which is numerically more stable and solver-friendly.
\begin{equation}
\sum_{m \in \mathcal{M}_{ex}} \!\left(q^{\mathrm{sell}}_{m,t} - q^{\mathrm{buy}}_{m,t}\right)
- P^{\mathrm{dis}}_t
\;\le\;
P^{\mathrm{bat}}_{\max}\,(1 - b^{\mathrm{dis}}_t) + s^{\mathrm{dis}}_t
\label{eq:discharge_coupling}
\end{equation}

\begin{equation}
P^{\mathrm{dis}}_t -
\sum_{m \in \mathcal{M}_{ex}} \!\left(q^{\mathrm{sell}}_{m,t} - q^{\mathrm{buy}}_{m,t}\right)
\;\le\;
P^{\mathrm{bat}}_{\max}\,(1 - b^{\mathrm{dis}}_t) + s^{\mathrm{dis}}_t
\label{eq:discharge_coupling_b}
\end{equation}

\begin{equation}
P^{\mathrm{dis}}_t + s^{\mathrm{dis}}_t
\;\ge\;
\sum_{m \in \mathcal{M}_{ex}} \!\left(q^{\mathrm{sell}}_{m,t} - q^{\mathrm{buy}}_{m,t}\right)
\label{eq:discharge_binary_coupling}
\end{equation}

\begin{equation}
\sum_{m \in \mathcal{M}_{ex}} \!\left(q^{\mathrm{sell}}_{m,t} + q^{\mathrm{buy}}_{m,t}\right)
- P^{\mathrm{ch}}_t
\;\ge\;
P^{\mathrm{bat}}_{\max}\,(1 - b^{\mathrm{ch}}_t) + s^{\mathrm{ch}}_t
\label{eq:charge_coupling}
\end{equation}

\begin{equation}
P^{\mathrm{ch}}_t +
\sum_{m \in \mathcal{M}_{ex}} \!\left(q^{\mathrm{sell}}_{m,t} - q^{\mathrm{buy}}_{m,t}\right)
\;\le\;
P^{\mathrm{bat}}_{\max}\,(1 - b^{\mathrm{ch}}_t) + s^{\mathrm{ch}}_t
\label{eq:charge_coupling_b}
\end{equation}

\begin{equation}
P^{\mathrm{ch}}_t + s^{\mathrm{ch}}_t
\;\ge\;
\sum_{m \in \mathcal{M}_{ex}} \!\left(q^{\mathrm{buy}}_{m,t} - q^{\mathrm{sell}}_{m,t}\right)
\label{eq:charge_binary_coupling}
\end{equation}

% \begin{equation}
% \forall\, t \in \mathcal{T} \nonumber
% \end{equation}

In addition to the global charging and discharging limits, each individual market $m \in \mathcal{M}_{ex}$ is subject to its own feasible trading range at every time step~$t$. 
%\paragraph{Market-Specific Trading Limits.}
Equations~\eqref{eq:market_buy_upper_limit} and~\eqref{eq:market_sell_upper_limit} ensure that the power bought or sold in a single market does not exceed the residual available charging or discharging capacity after accounting for concurrent balancing reserve commitments. 
The reduction in available power is quantified by the blocked power factor~$\phi_m$ [–], which specifies the additional MW that must remain available per MW of reserve capacity committed in market~$m$.
\begin{equation}
q^{\mathrm{buy}}_{m,t} 
\le P^{\mathrm{bat}}_{\max} - 
\sum_{r \in \mathcal{M}_{neg}} \left( \phi_r \cdot P^{\mathrm{res}}_{r,t} \right)
% , \quad \forall m \in \mathcal{M}_{ex},\; t \in \mathcal{T}
\label{eq:market_buy_upper_limit}
\end{equation}

\begin{equation}
q^{\mathrm{sell}}_{m,t} 
\le P^{\mathrm{bat}}_{\max} - 
\sum_{r \in \mathcal{M}_{pos}} \left( \phi_r \cdot P^{\mathrm{res}}_{r,t} \right)
%, \quad \forall m \in \mathcal{M}_{ex},\; t \in \mathcal{T}
\label{eq:market_sell_upper_limit}
\end{equation}

\subsection{Sequential Market Optimization}

Let $\mathcal{M} = \{ m_{FCR}, m_{aFRR,cap}, \dots, m_{IDA_1} \}$ be the set of markets, ordered by their respective gate-closure times $t_{\mathrm{GCT}}^{m}$, compare Fig.~\ref{fig:Ablauf}.
For each market $m \in \mathcal{M}$, the following sequential dispatch optimization problem is solved:
\begin{equation}
\begin{aligned}
\max_{x^{m}} \quad & \Pi^{m}\!\left(x^{m}, \hat{c}^{m}_{t_{\mathrm{GCT}}^{m}}\right) \\
\text{s.t.} \quad
& g(x^{m}) \le 0, \\
& h\!\left(x^{m}, \mathrm{SoC}_{\mathrm{in}}^{m}\right) = 0, \\
& \mathrm{SoC}_{\mathrm{out}}^{m} = f\!\left(x^{m}, \mathrm{SoC}_{\mathrm{in}}^{m}\right),
\end{aligned}
\label{eq:seq_opt_market}
\end{equation}
where  
\begin{itemize}
    \item $x^{m}$ denotes the decision variables (e.g., $P^{\mathrm{res}}_{m,t}$, $q^{\mathrm{buy/sell}}_{m,t}$, $P^{\mathrm{ch}}_t$, $P^{\mathrm{dis}}_t$),
    \item $\hat{c}^{m}_{t_{\mathrm{GCT}}^{m}}$ is the (forecasted) price vector, which considers information from already cleared markets at \ac{GCT}
    \item $\Pi^{m}(\cdot)$ denotes the expected profit function of market $m$,
    \item $g(\cdot)$ and $h(\cdot)$ represent inequality and equality constraints as defined in Section~\ref{sec:modeling},
    \item $\mathrm{SoC}_{\mathrm{in}}^{m}$ and $\mathrm{SoC}_{\mathrm{out}}^{m}$ are the states of charge before and after market $m$.
\end{itemize}

After solving for market $m_k$, the resulting $\mathrm{SoC}_{\mathrm{out}}^{m_k}$ becomes a fixed boundary condition for the subsequent iteration:
\begin{equation}
\mathrm{SoC}_{\mathrm{in}}^{m_{k+1}} = \mathrm{SoC}_{\mathrm{out}}^{m_k}.
\label{eq:seq_soc_link}
\end{equation}

\paragraph*{Residual capacity optimization}
A subsequent optimization is performed for the remaining markets $\mathcal{M}_{>k} = \{ m_{k+1}, \dots, m_N \}$ in order to determine the maximum feasible bid quantity and opportunity-based minimum bid price for $m_k$:
\begin{equation}
\begin{aligned}
\max_{x^{\mathcal{M}_{>k}}} \quad & U^{m_k}\!\left(x^{\mathcal{M}_{>k}}\right) \\
\text{s.t.} \quad 
& \mathrm{SoC}_{\mathrm{in}}^{m_{k+1}} = \mathrm{SoC}_{\mathrm{out}}^{m_k}, \\
& g\!\left(x^{\mathcal{M}_{>k}}\right) \le 0,
\end{aligned}
\label{eq:seq_opt_residual}
\end{equation}
where $U^{m_k}(\cdot)$ denotes the utilization function, yielding the maximum bid quantity $L_b$ (cf.~\eqref{eq:opp_block_quantity}) and the minimum bid price $\pi_b$ (cf.~\eqref{eq:min_bid_price}) for the bidding strategy.

\paragraph{Allocation and fixation.}
After simulated allocation and settlement (see Section~\ref{subsubsec:opportunitycostscalc}), the awarded decision variables of $m_k$ become fixed parameters for all subsequent optimizations:
\begin{equation}
x^{m_k}_t = x^{m_k, \mathrm{alloc}}_t,
\quad \forall\, t \in \mathcal{T},\; m > m_k.
\label{eq:seq_fixation}
\end{equation}
Thus, previously cleared market results are preserved, while subsequent markets are optimized under updated boundary and forecast information corresponding to their respective \ac{GCT}.

\subsubsection{Objective Functions}
\label{subsec:objective}

The model supports several objective variants that follow the goal of maximizing expected market revenues and capacity payments while penalizing imbalance between the market schedule and the battery operation plan, and optionally taking degradation costs or other battery-related restrictions into account.
The generic objective reads
\begin{align}
\max\; \Pi \;=\;&
\underbrace{\sum_{t\in\mathcal{T}} \sum_{m\in\mathcal{M}_{ex}}
c^{\mathrm{ex}}_{m,t}\,\bigl(q^{\mathrm{sell}}_{m,t}-q^{\mathrm{buy}}_{m,t}\bigr)}_{\text{energy market revenue}}
\nonumber\\[0.5em]
&+\;
\underbrace{\sum_{t\in\mathcal{T}} \sum_{m\in\mathcal{M}_{cap}}
c^{\mathrm{cap}}_{m,t}\, P^{\mathrm{res}}_{m,t}}_{\text{capacity payments}}
\nonumber\\[0.5em]
&-\;
\underbrace{\sum_{t\in\mathcal{T}}\!\Bigl(
\kappa^{\mathrm{dis}}\, s^{\mathrm{dis}}_{t} + \kappa^{\mathrm{ch}}\, s^{\mathrm{ch}}_{t}\Bigr)}_{\text{imbalance penalties}}
% \;-\;
% \underbrace{J^{\mathrm{deg}}}_{\text{(optional) battery penalties}}.
% _{q^{\mathrm{sell}}_{m,t},q^{\mathrm{buy}}_{m,t},P^{\mathrm{res}}_{m,t},s^{\mathrm{dis}}_{t},s^{\mathrm{ch}}_{t}}
\label{eq:zielfunktion}
\end{align}

Different objective modes are realized by plugging different price series into
\eqref{eq:zielfunktion}. 
In the standard case, the median of the respective forecasts is assumed for $c^{\mathrm{ex}}_{m,t}$ and $c^{\mathrm{cap}}_{m,t}$. However, it is also possible, for example, to work with perfect foresight, i.e., the actual prices. $c^{\mathrm{ex}}_{m,t}$ and $c^{\mathrm{cap}}_{m,t}$ are to be understood as €/MW scaled to $\Delta t$. In the sequential model, volume-weighted prices are used instead of forecasts after successful allocation if the market has already been allocated.

The maximum possible bids are determined in each round. Therefore, besides the revenue–maximizing objective~\eqref{eq:zielfunktion}, the model supports
auxiliary objective variants that push a single product to its feasible maximum
under all constraints (market blocks, \ac{SOC} bounds due to balancing reserves, charge/discharge limits, etc.). These modes are used, e.g., to derive bid quantities during the
sequential day-ahead workflow. Concretely,

\begin{align}
\max\;& \sum_{t\in\mathcal{T}} P^{\mathrm{res}}_{\mathrm{FCR},t}
, \label{eq:max_fcr}\\
\max\;& \sum_{t\in\mathcal{T}} \Bigl(P^{\mathrm{res}}_{\mathrm{aFRR\_pos},t}
      + P^{\mathrm{res}}_{\mathrm{aFRR\_neg},t}\Bigr)
, \label{eq:max_afrr}\\
\max\;& \sum_{t\in\mathcal{T}} q^{\mathrm{buy}}_{\mathrm{DAA},t}
\quad\text{or}\quad
\max\; \sum_{t\in\mathcal{T}} q^{\mathrm{sell}}_{\mathrm{DAA},t}
 \label{eq:max_daa}
\end{align}

These objectives maximize the power scheduled in the respective product (MW)
over all time steps, without price weights. They are evaluated with regard to the full constraint set from Section~\ref {sec:modeling}, so resulting sums represent feasible upper bounds for bidding in the corresponding market while respecting battery physics and concurrent reserve commitments.

%\subsubsection{risk-neutral Optimization}
%The overall optimization in the risk-neutral version is shown in Formula \ref{eq:objective_function_revenue}. The 0.5 quantiles of the price forecasts are used for all markets.

%\begin{equation}
%\begin{aligned}
%\max \quad \sum_{t=1}^{T} \Bigg\{ 
%    &\underbrace{
%        \tilde{p}^{\mathrm{cap}}_{\mathrm{FCR},t} \cdot q^{\mathrm{cap}}_{\mathrm{FCR},t}
%    }_{\text{FCR}} \\[0.5em]
%    &+
%    \underbrace{
%        \left( 
%            0.5 \cdot \tilde{p}^{\mathrm{clr}}_{\mathrm{aFRR+},t}
%            + 0.5 \cdot \tilde{p}^{\mathrm{avg}}_{\mathrm{aFRR+},t}
%        \right) \cdot q^{\mathrm{cap}}_{\mathrm{aFRR+},t}
%    }_{\text{aFRR+}} \\[0.5em]
%    &+
%    \underbrace{
%        \left( 
%            0.5 \cdot \tilde{p}^{\mathrm{clr}}_{\mathrm{aFRR-},t}
%            + 0.5 \cdot \tilde{p}^{\mathrm{avg}}_{\mathrm{aFRR-},t}
%        \right) \cdot q^{\mathrm{cap}}_{\mathrm{aFRR-},t}
%    }_{\text{aFRR-}} \\[0.5em]
%    &+
%    \underbrace{
%        \tilde{p}^{\mathrm{ex}}_{\mathrm{DAA},t} \cdot \left( %q^{\mathrm{sell}}_{DAA,t} - q^{\mathrm{buy}}_{DAA,t} \right)
%    }_{\text{DAA}} \\[0.5em]
%    &+
%    \underbrace{
%        \tilde{p}^{\mathrm{ex}}_{\mathrm{IDA1},t} \cdot \left( %q^{\mathrm{sell}}_{IDA1,t} - q^{\mathrm{buy}}_{IDA1,t} \right)
%    }_{\text{IDA1}}
%\Bigg\}
%\end{aligned}
%\label{eq:objective_function_revenue}
%\end{equation}

The maximum bid volume can now be determined. The calculation of the minimum price based on \acp{OC} will be considered on a market-specific basis.

\subsection{Bidding Heuristic}
\label{subsubsec:opportunitycostscalc}
For all products/markets, the \ac{OC} calculation follows the same general multi-stage structure.  Let $m_c$ denote the current market under consideration and $\mathcal{B}_{m_c}$ its block set.

First, the complete overall-optimization problem is solved using the revenue–maximizing objective~\eqref{eq:zielfunktion} or in its modified form. The resulting battery \ac{SOC} trajectory $\mathrm{SoC}^{(1)}_t$ and the market-specific allocations $P^{\mathrm{res}(1)}_{m,t}$ or $q^{\mathrm{buy/sell}(1)}_{m,t}$ form the reference.
Then, the \ac{SOC} is fixed at the end of each block $b \in \mathcal{B}_{m_c}$ to the first baseline value,  

\begin{equation}
\mathrm{SoC}^{(2)}_{t_b} = \mathrm{SoC}^{(1)}_{t_b}
\quad \forall\, t_b \in \mathcal{B}_{m_c}
\label{eq:opp_fix_soc}
\end{equation}

and all variables of the current market $m_c$ are set to zero:
\begin{equation}
P^{\mathrm{res}(2)}_{m_c,t} = 0
\quad\text{or}\quad
q^{\mathrm{buy/sell}(2)}_{m_c,t} = 0,
\quad \forall\, t\in\mathcal{T}.
\label{eq:opp_deactivate_mc}
\end{equation}
It is assumed here that, due to degradation costs, the battery will generally remain in \ac{SOC} ranges that allow it to continue participating in the balancing reserve in the subsequent block. For simplicity, the \ac{SOC} is fixed to take into account the energy value. However, to ensure that the problem can still be solved without resorting to an expensive imbalance energy, the option of deactivating the constraints \eqref{eq:buy_step_constraint} and \eqref{eq:sell_step_constraint} is selected, as mentioned above. This means that the charge value is determined on the basis of market prices.
Reoptimizing \eqref{eq:zielfunktion} under these restrictions yields the objective contribution $V_b$ of the remaining markets in block $b$.

\begin{equation}
\begin{split}
V_b
= \sum_{t\in b}
\Biggl[
  \sum_{m\in\mathcal{M}_{ex}}
  c^{\mathrm{ex}}_{m,t}\!\left(q^{\mathrm{sell}(2)}_{m,t}-q^{\mathrm{buy}(2)}_{m,t}\right) \\
  + \sum_{m\in\mathcal{M}_{cap}}
  c^{\mathrm{cap}}_{m,t} P^{\mathrm{res}(2)}_{m,t}
\Biggr]
\label{eq:opp_block_value}
\end{split}
\end{equation}

As seen in \eqref{eq:opp_block_value}, slack variables are completely excluded to ensure calculation of \ac{OC} based on market values.   

A separate problem is solved with the objective of maximizing the scheduled quantity of the current market (cf. \eqref{eq:max_fcr}–\eqref{eq:max_daa}), producing the feasible upper limit per time step. The maximum bid quantity of the block $L_b$ is the minimum in the block. For energy exchange markets, the initial optimization continues to determine whether the market buys or sells in a time block. The variable that is 0 is then fixed at 0. This means that for each time step $t$, one variable is maximized while the other is 0.

\begin{equation}
L_b =
\min_{t\in b}
\begin{cases}
P^{\mathrm{res}}_{m_c,t}, & m_c\in\mathcal{M}_{cap},\\[0.15em]
\bigl|q^{\mathrm{sell}}_{m_c,t}-q^{\mathrm{buy}}_{m_c,t}\bigr|, & m_c\in\mathcal{M}_{ex}
\end{cases}
\label{eq:opp_block_quantity}
\end{equation}

The \ac{OC} for a block $b$ reflects the reduction in the overall objective value caused by reserving $L_b$ for the current market $m_c$.  
For products that require to block parts of battery power (e.g., FCR, aFRR), this reservation reduces the amount of battery power available for other markets.  
To account for this, the usable share of the battery’s nominal power $P_{\max}^{\mathrm{bat}}$ is reduced by the blocking factor $\phi_{m_c}$, which represents the fraction of $L_b$ that effectively blocks subsequent market participation due to technical or regulatory requirements.

Since markets have different minimum bid increments, remaining usable capacity is rounded down to the smallest bid increment $\min(\Delta q_{c+i})$ of all subsequent markets $m_{c+i}$.  
This ensures that \ac{OC} reflects complete power levels that can not be traded in later stages.

The resulting loss–profit share $\rho_b$ expresses the fraction of potential capacity that is not available for subsequent markets due to the block reservation:

\begin{equation}
\rho_b = 
\frac{
\left\lfloor 
\frac{\max\left\{0,\, P_{\max}^{\mathrm{bat}} - \phi_{m_c} L_b\right\}}
{\min(\Delta q_{c+i})}
\right\rfloor
}{
\frac{P_{\max}^{\mathrm{bat}}}{\min(\Delta q_{c+i})}
}.
\label{eq:loss_profit_share}
\end{equation}
A value of $\rho_b = 1$ means that block reservation has no effect on available battery power for later markets, whereas $\rho_b = 0$ indicates that reservation eliminates usable capacity.

The loss–profit share $\rho_b$ from \eqref{eq:loss_profit_share} thus quantifies the relative share of the battery's nominal capacity that remains usable for subsequent markets after reserving $L_b$ for $m_c$.  
This factor directly scales the expected value $V_b$ of the remaining markets in block $b$: if $\rho_b$ is reduced, a proportionally larger part of the potential profit is lost.

The \ac{OC} for block $b$ is then obtained as the foregone share of the block’s value:
\begin{equation}
\mathrm{OC}_b = \left( 1 - \rho_b \right) \cdot V_b,
\label{eq:opp_cost_block}
\end{equation}
where $V_b$ is the optimal objective value of all markets except $m_c$, subject to the \ac{SOC} and deactivation constraints \eqref{eq:opp_fix_soc}–\eqref{eq:opp_deactivate_mc}, $\rho_b$ reflects the capacity that remains tradable for subsequent products after reserving $L_b$, and $(1 - \rho_b)$ is the share of potential profit lost due to the reservation.

% By expressing $\mathrm{OC}_b$ in this way, the calculation links technical constraints with their economic impact.  
% In particular, balancing reserves, tend to cause higher \acp{OC} because their blocking factor $\phi_{m_c}$ significantly reduces the available battery capacity for other markets.

Finally, the minimum bid price $\pi_b$ for block $b$ is determined by distributing \acp{OC} \eqref{eq:opp_cost_block} over offered quantity and temporal extent of the block.  
Let $B_{m_c}$ denote the number of time steps in a single block of market $m_c$, and $\Delta t$ the step duration in hours. The minimum bid price reads as

\begin{equation}
\pi_b =
\begin{cases}
\frac{\mathrm{OC}_b}{L_b \cdot B_{m_c} \cdot \Delta t}, & \text{if } L_b > 0, \\[0.3em]
0, & \text{otherwise}.
\end{cases}
\label{eq:min_bid_price}
\end{equation}

This expression yields the break-even price below which participation in $m_c$ would reduce the overall expected profit of the multi-market portfolio.  
In practice, $\pi_b$ can be further adjusted by product-specific heuristics or market rules. Equation~\eqref{eq:min_bid_price} thus provides the economic floor for bid pricing, linking the technical capacity interactions of the storage system with the multi-market opportunity structure.

\subsubsection{Bidding Strategy for FCR}
In the FCR capacity market a pay–as–cleared auction is used. Consequently, there is no benefit from splitting the available volume into multiple prices. Therefore, the full feasible block volume is bidded at a single price that covers the block’s \acp{OC}.

Let $m_c=\mathrm{FCR}$ and let $b\in\mathcal{B}_{\mathrm{FCR}}$ denote a 4-hour block. From quantity maximization (cf.~\eqref{eq:max_fcr}) the feasible block volume $L_b$ is obtained. The minimum bid price follows from the opportunity–cost calculation \eqref{eq:min_bid_price} and is rounded up to the nearest possible monetary amount, i.e., to the nearest cent. The submitted FCR bid for block $b$ is thus
\begin{equation}
q^{\mathrm{cap}}_{\mathrm{FCR},b} = L_b,
\quad
g^{\mathrm{bid}}_{\mathrm{FCR},b} = \bigl\lceil 100 \cdot \pi_b \bigr\rceil / 100,
\label{eq:fcr_single_bid}
\end{equation}
i.e., the entire admissible quantity $L_b$ is offered on the (rounded) opportunity-cost $\pi_b$.

Since remuneration is based on the pay-as-cleared procedure, ladder bids would not improve expected revenue; hence \eqref{eq:fcr_single_bid} suffices. Bidding at true value is a weakly dominant strategy here in terms of game theory.

\subsubsection{Bidding Strategy for aFRR Capacity}

Unlike the \ac{FCR} market, which is settled pay-as-cleared, the aFRR capacity market applies a pay-as-bid settlement. 
This makes the pricing of each individual bid critical. The offering of all available capacity $L_b$ at a single price would expose the operator to the risk of partial or complete rejection in volatile price periods.  
A more resilient approach is to split the available block volume into several smaller bids and price them along a spectrum between conservative bids with high probability of acceptance and aggressive ones with higher potential margins.

For a given product block $b$ and direction $r\in\{+,-\}$, the feasible bid set $L_b^r$ is determined by simultaneously maximizing the positive and negative aFRR sets as described in \eqref{eq:max_afrr}. Since the problem can thus have multiple solutions, as a lexicographic tiebreak for \eqref{eq:max_afrr}, block-wise symmetry between
positive and negative aFRR is enforced by minimizing the absolute block
imbalance:

\begin{align}
\min\;& d_b
\label{eq:afrr_sym}\\
\text{s.t.}\quad
& d_b \ge
  \sum_{t\in\mathcal{T}_b}\!\bigl(
    P^{\text{res}}_{\text{aFRR\_neg},t}
    - P^{\text{res}}_{\text{aFRR\_pos},t}
  \bigr),\\
& d_b \ge
  \sum_{t\in\mathcal{T}_b}\!\bigl(
    P^{\text{res}}_{\text{aFRR\_pos},t}
    - P^{\text{res}}_{\text{aFRR\_neg},t}
  \bigr).
\end{align}

where $\mathcal{T}_b$ denotes the set of time steps in block $b$.

Given the minimum bid size $\Delta q_{m_c}$ prescribed by the market rules, the maximum number of bids is
\begin{equation}
n_b^r = \left\lfloor \frac{L_b^r}{\Delta q_{m_c}} \right\rfloor.
\label{eq:afrr_num_bids}
\end{equation}
This ensures that the full available capacity is offered in discrete, market-compliant steps.  
The quantity of blocks is then evenly distributed among these $n_b^r$ bids, with all but the first bid having exactly $\Delta q_{m_c}$ and the first absorbing any remainder.
\begin{equation}
b_{b,1}^r = L_b^r - (n_b^r-1)\,\Delta q_{m_c}, 
\quad
b_{b,i}^r = \Delta q_{m_c} \quad\text{for}\quad i=2,\dots,n_b^r.
\label{eq:afrr_bid_size}
\end{equation}
In Germany, the bid size is always 1 MW for optimal distribution of bids; in other countries, different market rules apply in some cases.
While bid sizes are therefore determined directly from the granularity of the market, bid prices require more strategic consideration.  
Starting with the minimum bid price $\pi_b$, which is derived from the \ac{OC} calculation \eqref{eq:min_bid_price} and represents the economic lower limit below which bids are unprofitable. 
Because operators may wish to temper the strictness of this floor, a risk factor $\theta\in(0,1]$ (default $\theta=0.5$) is introduced.  
This scales the opportunity-cost floor to an \emph{effective} minimum:
\begin{equation}
\pi_{b,\mathrm{eff}} = \theta \,\pi_b.
\label{eq:afrr_theta}
\end{equation}
A value of $\theta=0.5$ means that half of the opportunity is carried by positive direction and the other by negative direction. This is because the opportunity for the joint deactivation of aFRR is determined and it is necessary due to the simultaneity of the allocation. The effective floor is then combined with market forecasts to determine the final bid prices.  
Let $c_{pi,b}^r$ be the point forecast for clearing price in block $b$ and $c_{aver,b}^r$ the forecast for average awarded price.  
Each bid $i$ is placed on a convex combination of these two forecasts and the adjusted floor:

\begin{align}
g_{b,i}^r
&= \frac{i}{n_b^r+1}\,
   \max\{c_{pi,b}^r,\,\pi_{b,\mathrm{eff}},\,0\} \nonumber\\
&\quad + \frac{n_b^r+1-i}{n_b^r+1}\,
   \max\{c_{aver,b}^r,\,\pi_{b,\mathrm{eff}},\,0\}.
\label{eq:afrr_bid_price}
\end{align}

Low-index bids are thus priced closer to $c_{aver,b}^r$, maximizing the likelihood of acceptance, 
while high-index bids lean more towards $c_{pi,b}^r$, capturing additional margin potential if prices spike.  
This stepped pricing approach effectively creates a price–probability gradient across the offered capacity.

The procedure \eqref{eq:afrr_num_bids}–\eqref{eq:afrr_bid_price} is applied independently for $r=+$ and $r=-$, 
with direction-specific quantities $L_b^{+}$, $L_b^{-}$ and forecasts $c_{pi,b}^{+}, c_{pi,b}^{-}, c_{aver,b}^{+}, c_{aver,b}^{-}$.  
In this way, asymmetries in system conditions and market expectations between positive and negative aFRR are reflected directly in the bidding structure.

Finally, in accordance with market convention, all bid prices are rounded \emph{upwards} to the nearest cent to avoid underbidding relative to the intended price level:
\begin{equation}
\tilde{g}_{b,i}^r = \left\lceil 100 \cdot g_{b,i}^r \right\rceil / 100.
\end{equation}
This ensures that even after rounding, the submitted price does not fall below the calculated value.

\subsubsection{Bidding Strategy for DAA}
The Day-Ahead Auction market differs fundamentally from both the FCR and aFRR capacity markets in two key respects. Firstly, it is an energy market and not a capacity market, and secondly, the optimization problem is inherently directional, as buy and sell bids are submitted simultaneously. 
The bidding strategy focuses on selecting the optimal direction and pricing the corresponding energy volume.

First, a baseline model with objective function \eqref{eq:zielfunktion} and second, an \ac{OC} run are performed, as is known. The third problem is divided into an auxiliary problem that determines maximum possible volume of buy per time step and another auxiliary problem for maximum possible sale volume.  

From the overall-optimization run, trading direction of block $b$ is first determined.  
Let baseline net block volume be
\begin{equation}
L_b^{\mathrm{base}}
\;=\;
\min_{t\in b}\,
\bigl|\,q^{\mathrm{sell}(1)}_{\mathrm{DAA},t}-q^{\mathrm{buy}(1)}_{\mathrm{DAA},t}\,\bigr|,
\label{eq:daa_baseline_block}
\end{equation}

\begin{equation}
    s_b \;=\; \operatorname{sign}\!\left(
\sum_{t\in b} \bigl[q^{\mathrm{sell}(1)}_{\mathrm{DAA},t}-q^{\mathrm{buy}(1)}_{\mathrm{DAA},t}\bigr]\right)\in\{-1,+1\},
\end{equation}

where $s_b=+1$ indicates a \emph{sell} block and $s_b=-1$ a \emph{buy} block.  
In parallel, the two maximum-volume problems provide direction-specific upper bounds:
\begin{equation}
\bar L_b^{\mathrm{sell}}
\;=\;
\min_{t\in b}\,\Bigl(q^{\mathrm{sell}(4)}_{\mathrm{DAA},t}-q^{\mathrm{buy}(4)}_{\mathrm{DAA},t}\Bigr),
\label{eq:daa_ubs}
\end{equation}

\begin{equation}
    \bar L_b^{\mathrm{buy}}
\;=\;
\min_{t\in b}\,\Bigl(q^{\mathrm{buy}(3)}_{\mathrm{DAA},t}-q^{\mathrm{sell}(3)}_{\mathrm{DAA},t}\Bigr),
\end{equation}

so that direction-specific, \emph{capped} block quantity is given by

\begin{equation}
L_b^{\star}
\;=\;
\begin{cases}
\min\{L_b^{\mathrm{base}},\,\bar L_b^{\mathrm{sell}}\}, & s_b=+1,\\[0.3em]
\min\{L_b^{\mathrm{base}},\,\bar L_b^{\mathrm{buy}}\}, & s_b=-1,
\end{cases}
\label{eq:daa_capped}
\end{equation}

with sign $s_b$ representing the trading direction.

Physical and economic consistency across blocks is ensured by fixing the \ac{SOC} values of the baseline run at all block boundaries, while in the opportunity-cost run the DAA variables are set to zero.  
From this \emph{opportunity problem}, the block value $V_b$ is obtained according to \eqref{eq:opp_block_value}.  
The corresponding opportunity-cost floor price per MW and block is then
\begin{equation}
\pi_b
\;=\;
\frac{V_b}{\bigl|L_b^{\star}\bigr|\; B_{\mathrm{DAA}}\;\Delta t}
\label{eq:daa_block_price_floor}
\end{equation}
where $B_{\mathrm{DAA}}$ is the number of time steps in the DAA block and $\Delta t$ is the step duration.

Market granularity is respected when setting the bid quantity.  
With the DAA increment $\Delta q_{\mathrm{DAA}}=\SI{0.1}{MW}$, the quantized block volume is
\begin{equation}
\tilde L_b
\;=\;
s_b \cdot \Delta q_{\mathrm{DAA}}
\cdot
\left\lfloor
\frac{\bigl|L_b^{\star}\bigr|}{\Delta q_{\mathrm{DAA}}}
\right\rfloor,
\label{eq:daa_quantization}
\end{equation}
which is submitted as a single block bid.  
The associated bid price uses the cost floor \eqref{eq:daa_block_price_floor} and is rounded to the nearest cent, \emph{directionally} according to the trade:  
for \emph{sell} blocks, prices are rounded \emph{up} to ensure they are above the \ac{OC},  
for \emph{buy} blocks, prices are rounded \emph{down} (with a negative sign) to ensure they remain below:
\begin{equation}
g_b \;=\;
\begin{cases}
\displaystyle \left\lceil 100\,\pi_b \right\rceil / 100, & s_b=+1 \ \text{(Sell)},\\[0.6em]
\displaystyle -\,\left\lfloor 100\,\pi_b \right\rfloor / 100, & s_b=-1 \ \text{(Buy)}.
\end{cases}
\label{eq:daa_rounding}
\end{equation}

In the DAA case, the loss-profit share $\rho_b$ from \eqref{eq:loss_profit_share} is not necessary, as the reserved capacity effectively prevents other market allocations from subsequent energy exchange markets in the same block, unless arbitrage is traded between several follow-up markets. Consequently, $\rho_b = 0$ for all blocks.

The block direction comes from the baseline schedule, the quantity is safeguarded by the respective maximum-volume auxiliary models and quantized to market increments,  
and the block price reflects the \ac{OC} per MW and block duration, with cent-accurate, direction-specific rounding.

\subsubsection{Bidding Strategy for IDA 1}  
In the intraday auction considered here, the market is assumed to be the last one traded in the sequential optimization process.  
As there are no subsequent markets whose allocations could be affected by the IDA, no cross-market \acp{OC} need to be considered.  
All relevant information for bid construction is thus taken directly from the overall optimization.  
This also implies that the loss--profit share $\rho_b$ is always
\begin{equation}
\rho_b = 0,
\end{equation}
since there is no following market.

First, the baseline market position is determined for each time step $t$, i.e.\ whether the storage asset is scheduled to sell, buy, or remain neutral.  
The $IDA_i$ market is cleared in fixed product blocks of duration $B_{\mathrm{IDA_i}}$ (in quarter-hours).  
For each block $b$, the feasible block quantity $L_b$ is obtained from the baseline schedule by taking the minimum power across all time steps in the block:
\begin{equation}
L_b = \min_{t \in b} \left| q^{\mathrm{sell}}_{m_c,t} - q^{\mathrm{buy}}_{m_c,t} \right|,
\quad m_c = \text{IDA}_i.
\label{eq:ida_block_quantity}
\end{equation}
This ensures delivery of offered quantity throughout the block.

Since there are no follow-up markets, the block value $V_b$ is calculated directly as the sum of the objective contributions from all time steps in $b$ for the IDA market:
\begin{equation}
V_b = \sum_{t \in b} \mathrm{ObjContr}_{m_c,t}.
\label{eq:ida_block_value}
\end{equation}
The economic floor price for the bid is then obtained by distributing the block value over the block’s energy volume:
\begin{equation}
\pi_b = \frac{V_b}{L_b \cdot B_{\mathrm{IDA}} \cdot \Delta t},
\label{eq:ida_floor_price}
\end{equation}
where $\Delta t$ is the time step length in hours.

To reflect the pay-as-cleared settlement mechanism, the bid price is rounded in a direction that favors acceptance. For sales ($L_b > 0$),  the price is round to the next cent:
    \begin{equation}
    g_b = \left\lfloor 100 \cdot \pi_b \right\rfloor / 100,
    \end{equation}
    slightly undercutting competing offers.
   For purchases ($L_b < 0$), round \emph{up} to the next cent:
    \begin{equation}
    g_b = -\left\lceil 100 \cdot \pi_b \right\rceil / 100,
    \end{equation}
    to avoid rejection when willing to pay above marginal price.

The resulting bid for block $b$ consists of signed block quantity $L_b$ (quantized to 0.1\,MW to match market granularity) and adjusted price $g_b$.  
Thus, each bid mirrors the baseline optimal schedule for IDA without further strategic adjustments, as no inter-market trade-off is relevant at this stage.

\subsection{Market Clearing and Settlement}
After bid construction, allocation is performed independently for each delivery timestamp~$\tau$.
Let $\mathcal{S}_\tau$ denote the set of submitted bids for~$\tau$, with bid $i\in\mathcal{S}_\tau$ characterized by a signed quantity $q_i$ [MW] and a price $g_i$ [€/MWh]. For energy trading markets, a positive $q_i>0$ denotes a \emph{sell} bid, a negative $q_i<0$ a \emph{buy} bid. For balancing capacity markets, the quantity is positive. The real market clearing price for timestamp~$\tau$ is called $\pi^{\mathrm{clr}}_\tau$.

Using $\operatorname{sgn}(q_i)\in\{-1,1\}$, the sign–aware acceptance test can be written compactly as
\begin{equation}
\text{bid $i$ accepted at }\tau
\;\Longleftrightarrow\;
\operatorname{sgn}(q_i)\, g_i \;\le\;\operatorname{sgn}(q_i)\,\pi^{\mathrm{clr}}_\tau.
\label{eq:alloc_accept_rule}
\end{equation}
Equivalently, in a product–agnostic form used,
\begin{equation}
g_i\, q_i \;\le\; \pi^{\mathrm{clr}}_\tau \, q_i,
\quad \forall\, i\in\mathcal{B}_\tau,
\label{eq:alloc_accept_rule_prod}
\end{equation}
which reduces to $g_i\le\pi^{\mathrm{clr}}_\tau$ for sells ($g_i>0$) and
$g_i\ge\pi^{\mathrm{clr}}_\tau$ for buys ($g_i<0$).

% If a per–timestamp allocation cap $Q^{\max}_\tau$ is enforced by the auction
% (e.g.\ network limits), the clearing respects
% \begin{equation}
% \sum_{i\in\mathcal{B}_\tau} \mathbb{1}_{\{\text{accepted}\}}\, g_i
% \;\le\; Q^{\max}_\tau,
% \label{eq:alloc_cap}
% \end{equation}
% with ties resolved by price–time priority.
% In the this work we consider the unconstrained case
% $Q^{\max}_\tau=+\infty$.

Given the pricing mode, the price paid per accepted bid is
\begin{equation}
s_i \;=\;
\begin{cases}
\pi^{\mathrm{clr}}_\tau, & \text{pay-as-cleared},\\[0.25em]
g_i, & \text{pay-as-bid},
\end{cases}
\quad \forall\, i \text{ accepted at } \tau.
\label{eq:settlement_rule}
\end{equation}
Rejected bids receive $s_i=0$. 

After each market has been cleared and allocations have been determined, the subsequent optimization is initiated accordingly. In particular, following the final day-ahead market allocation, the resulting portfolio may contain open positions, as not all submitted bids are necessarily accepted. This can render the problem infeasible without the use of imbalance energy. To address this, the optimization is repeated, incorporating bid increments and slack variables for imbalance energy in order to derive a feasible battery target schedule for the intraday optimization stage. Any remaining open positions must be covered by the operator on \ac{IDC} market, if necessary.

\section{Validation of sequential optimization of BESS}
The sequential optimization must be validated to ensure that the applied methodology delivers consistent and reliable results under realistic operating conditions. In this context, the optimization process is designed to update the dispatch schedule at several points during the day as new information from the rolling forecasts becomes available. Validation should test whether rolling forecasts at each market stage trigger updates and what deviations they cause, addressing the second part (ii) of the research question.

\subsection{Single day validation}
%Day-Opt. Validation
To conduct this validation, a sample day (April 1, 2024) is selected. The BESS is assumed to have a rated power of 3.65 MW and an energy capacity of 7.3 MWh. The efficiency is simplified as a constant of 0.95 for both input and output, to reduce computational complexity. For this day, the results of the individual optimization steps are analyzed and compared to identify possible changes between the consecutively executed time intervals.

\begin{figure}
	\centering
	\includegraphics[width=\columnwidth]{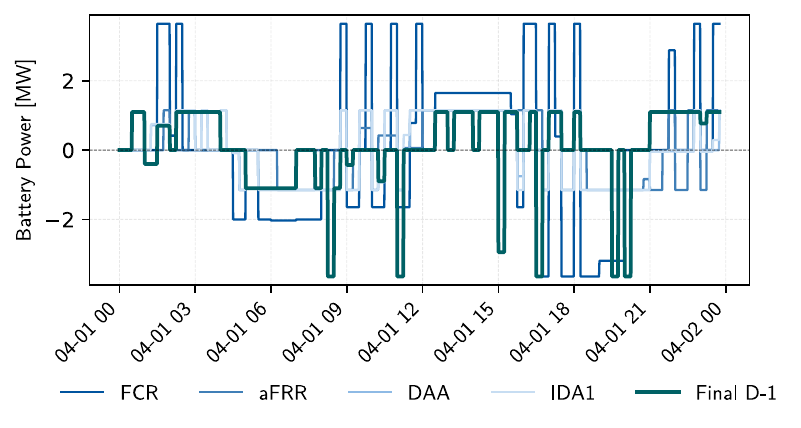}
	\caption{Comparison of D-1 power schedules at successive optimization stages.}
	\label{fig:p_fahrplan_d-1}
\end{figure}

\begin{figure}
	\centering
	\includegraphics[width=\columnwidth]{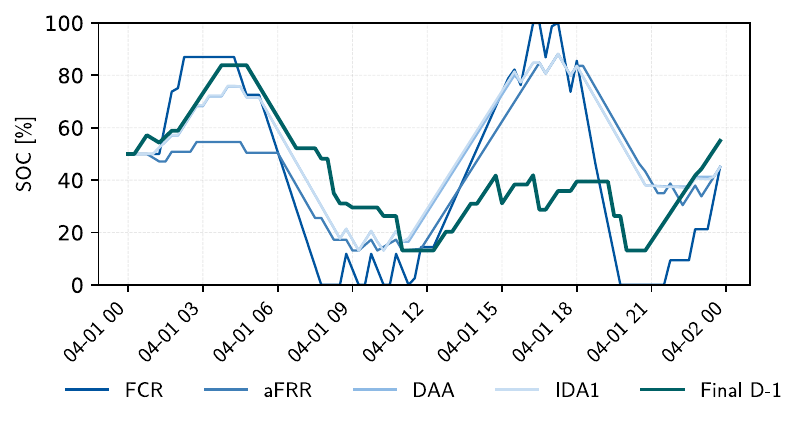}
	\caption{Comparison of state-of-charge trajectories during D-1 optimization.}
	\label{fig:soc_fahrplan_d-1}
\end{figure}

Figure \ref{fig:Ablauf} illustrates the chronological sequence of the optimization procedure. The results of the individual overall optimizations for April 1, 2024 (day D) are shown in Figure \ref{fig:p_fahrplan_d-1}. The schedule for Day D, which was generated at \textcolor{rwth}{8 a.m.}, can be seen, as well as the schedules for \textcolor{rwth-75}{9 a.m.}, \textcolor{rwth-50}{12 p.m.}, \textcolor{rwth-25}{3 p.m.}, and the schedule at the end of the day-ahead optimization process (\textcolor{petrol}{Final D-1}). The figures indicate the quarter-hour intervals during which the \ac{BESS} is scheduled to charge or discharge. The corresponding SoC trajectories at the different optimization stages are shown in Figure \ref{fig:soc_fahrplan_d-1}.

The schedule are changing between times due to the revised forecasts, the auctions that have already taken place, and the resulting allocations. For example, in the first time step, FCR is only planned for part of the day. The bid for FCR was the respective opportunity costs for the 4-hour blocks. The \ac{BESS} was allocated for all 6 product slices, which can be seen in the aFRR curve that no SoC states close to 100 \% or 0 \% are included in the schedule anymore, as this range is blocked for the allocated FCR.
The significant deviations from the final D-1 schedule to the schedule at 3 p.m. (IDA 1) can be explained by the fact that the model allows a tolerance for the final SoC in order to avoid having to plan with too much imbalance energy. The price for the imbalance schedule is the reBAP. The reBAP itself is not part of the optimization process and is not used for counter-optimization, as frequent use constitutes a breach of the balancing group contract with the TSO and therefore poses a legal risk for the provider. However, it is used later to determine the revenue from imbalance.

\begin{figure}
	\centering
	\includegraphics[width=\columnwidth]{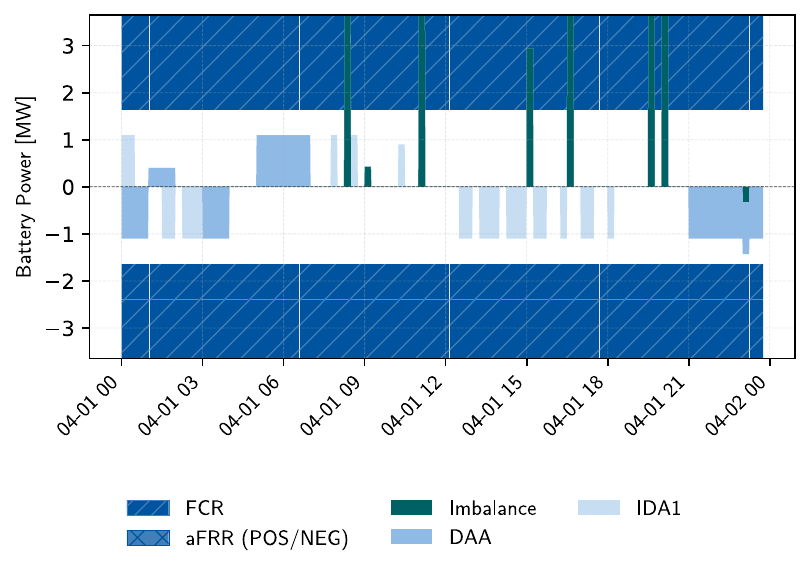}
	\caption{Allocated bids for D (final day-ahead schedule)}
	\label{fig:fahrplan_final_d1}
\end{figure}

Figure \ref{fig:fahrplan_final_d1} shows the final day-ahead schedule broken down by individual markets. The results indicate that the \ac{BESS} must provide FCR throughout day D while no capacity is reserved for aFRR. The trades from the DAA and IDA1 auctions are also shown. Trades that cannot be executed as bid due to regulatory requirements are balanced with imbalance energy in the day-ahead planning. This does not mean that the \ac{BESS} actually uses imbalance energy, as it still has the option to trade on the IDC after the day-ahead planning.

\begin{figure}
	\centering
	\includegraphics[width=\columnwidth]{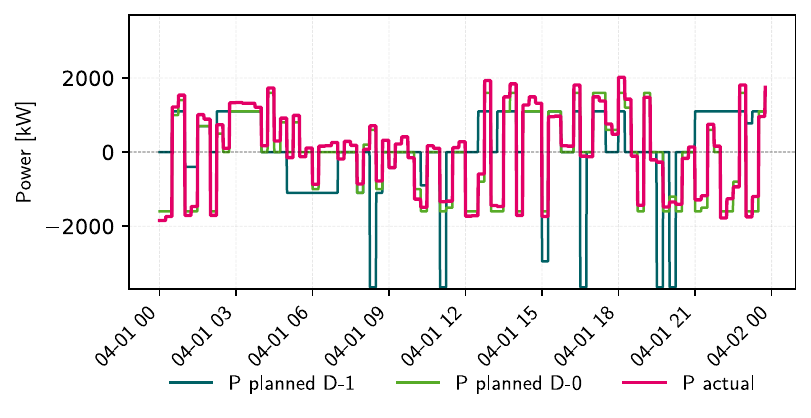}
	\caption{Comparison of planned and realized power dispatch across scheduling stages.}
	\label{fig:fahrplanvergleich_P}
\end{figure}

\begin{figure}
	\centering
	\includegraphics[width=\columnwidth]{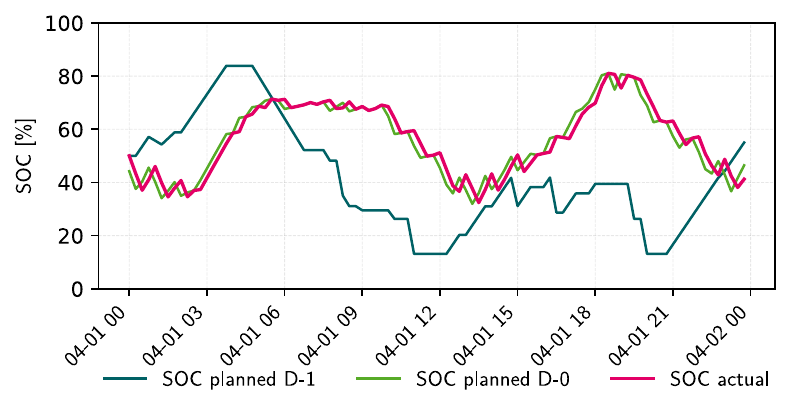}
	\caption{Comparison of planned and realized state-of-charge trajectories.}
	\label{fig:fahrplanvergleich_soc}
\end{figure}

The extent to which the planned dispatch from the previous day differs from the actual dispatch for April 1, 2024 can be seen in Figure \ref{fig:fahrplanvergleich_P}. Figure \ref{fig:fahrplanvergleich_soc} shows that the previous day’s schedule differs significantly more from the intraday schedule and the actual dispatch than these differ from each other. The closer the schedule is to dispatch time T, the more accurate it becomes.

In dispatch, it is no longer necessary to use imbalance energy, which was previously required in D-1 planning. Further changes between day-ahead planning and intraday planning can be explained by necessary storage management measures following FCR activations that already took place on day D, as these cannot be forecast in day-ahead planning and are included in intraday optimization as actual calls, as this is updated every 15 minutes. 
In addition, the IDC was not taken into account in the day-ahead planning. Active participation in the IDC takes place, and the result of this participation is included in the intraday optimization and the actual dispatch. Furthermore, it is possible to participate in FRR energy auctions that were not planned in D-1. However, this is not the case on the day in question, as the IDC was more profitable in the profit calculation than participation in aFRR energy. The differences between the intraday plan and dispatch can be caused by FCR provision.
Throughout the simulation horizon, the \ac{SOC} trajectory remains within the technical limits required for \ac{FCR} provision, and the maximum charging and discharging power of the \ac{BESS} is never exceeded in either the optimization results or the actual dispatch.
\section{Conclusion}
\label{sec:conclusion}

This work developed and validated a data-driven optimization framework for \ac{FTM} \ac{BESS} participating in power exchange and balancing service markets.

The proposed sequential optimization framework explicitly represents real market mechanisms, including \ac{GCT}, and bidding procedures.  
Feasible bid volumes and opportunity cost–based bid prices are derived at each market stage, enabling coordinated participation of the \ac{BESS} across multiple markets while maintaining operational feasibility.

Validation results confirm that the framework generates consistent and technically feasible schedules.  Sequential optimization steps remain coherent over the day and adapt effectively to updated forecasts and allocations.  
Intraday re-optimization substantially reduces deviations from actual dispatch compared to day-ahead plans, demonstrating the robustness of the approach under realistic information updates.

Addressing the research question posed in Sec.~\ref{subsec:contributions}, the findings confirm that a data-driven sequential optimization algorithm can realistically schedule BESS across markets by integrating rolling price forecasts, market-specific constraints, and opportunity-cost-based bidding. 
% This design enables market-consistent decision-making and improves operational alignment and revenue realization compared to single-stage day-ahead optimization.

Future work will extend the framework with \ac{RL}-based \ac{IDC} trading that executes the D–1 schedule and performs adaptive real-time trading under market uncertainty.

\bibliography{literature}

\end{document}